# Harnessing and control of optical rogue waves in supercontinuum generation

**John. M. Dudley,[1*] Goëry Genty[2] and Benjamin J. Eggleton[3]**

[1]Département d'Optique P. M. Duffieux, Institut FEMTO-ST, UMR 6174 CNRS-Université de Franche-Comté,

25030 Besançon, France

*Corresponding author: john.dudley@univ-fcomte.fr

[2]Tampere University of Technology,

Institute of Physics, Optics Laboratory,

FIN-33101 Tampere, Finland

[3]Centre for Ultra-high-Bandwidth Devices & Optical Systems (CUDOS),

School of Physics, University of Sydney, NSW 2006, Australia

## Abstract

We present a numerical study of the evolution dynamics of "optical rogue waves", statistically-rare extreme red-shifted soliton pulses arising from supercontinuum generation in photonic crystal fiber [D. R. Solli *et al.* Nature **450**, 1054-1058 (2007)]. Our specific aim is to use nonlinear Schrödinger equation simulations to identify ways in which the rogue wave dynamics can be actively controlled, and we demonstrate that rogue wave generation can be enhanced by an order of magnitude through a small modulation across the input pulse envelope and effectively suppressed through the use of a sliding frequency filter.

## 1. Introduction

Since its first observation by Ranka *et al.* in 2000 [1], supercontinuum (SC) generation in photonic crystal fiber has been the subject of extensive research [2–6]. Particular interest has focused on understanding the SC noise properties, as the SC amplitude and phase stability are key factors in assessing potential applications. Although initial research here concentrated on establishing guidelines for stable SC generation using femtosecond pulses [7–8], subsequent work has considered SC noise properties over a wider parameter range, from the picosecond to the continuous wave regime [9-11]. This has led to further studies of the various mechanisms by which input pulse noise is transferred to the output SC spectrum [12–14].

In this context, highly significant experiments have recently been reported by Solli *et al.* where a novel wavelength-to-time detection technique has allowed the direct characterization of the shot-to-shot statistics of a SC generated with picosecond pulses [15]. Although this regime of SC generation is well-known to exhibit fluctuations in the positions of Raman solitons on the SC long wavelength edge [6], Solli *et al.* have shown that these fluctuations contain a small number of statistically-rare "rogue" events associated with an enhanced red-shift and a greatly increased intensity. Crucially, because these experiments were performed in a regime where modulation instability (MI) plays a key role in the dynamics, it has been possible to propose an important correspondence with the hydrodynamic rogue waves of oceanic infamy [16], whose origin has also been discussed in terms of MI or, as it usually referred to in hydrodynamics, the Benjamin-Feir instability [17-19].

Although the links between optical and oceanic rogue waves will clearly require much further analysis, our objective here is to examine ways in which photonic technologies may be used to harness and control rogue wave generation in an optical context. Specifically, we use a generalized nonlinear Schrödinger equation model to study the evolution dynamics of optical rogue wave (or rogue soliton) generation, and we apply these results to determine conditions under which rogue soliton formation can be manipulated in a controlled way. By performing multiple simulations in the presence of noise, we examine the effect of input pulse modulation and spectral filtering on the SC generation dynamics, and we show that modifying the rogue wave generation process should indeed be possible using readily-available experimental techniques. Specifically, we demonstrate that rogue wave generation can be enhanced by an order of magnitude through a small modulation across the input pulse envelope and effectively suppressed through the use of a sliding frequency filter.

## 2. Numerical model and general features

Our simulations are based on the generalized nonlinear Schrödinger equation [6]:

$$\frac{\partial A}{\partial z} - \sum_{k \geq 2} \frac{i^{k+1}}{k!} \beta_k \frac{\partial^k A}{\partial t^k} = i\gamma \left(1 + i\tau_{shock} \frac{\partial}{\partial t}\right) \left(A(z,t) \int_{-\infty}^{+\infty} R(t') |A(z,t-t')|^2 \, dt'\right). \quad (1)$$

Here $A(z,t)$ is the field envelope and the $\beta_k$'s and $\gamma$ are the usual dispersion and nonlinear coefficients. The nonlinear response $R(t) = (1-f_R)\delta(t) + f_R h_R(t)$ includes instantaneous and Raman contributions. We use $f_R = 0.18$ and $h_R$ determined from the experimental fused silica Raman cross-section [6]. The self-steepening timescale $\tau_{shock}$ includes the dispersion of the nonlinearity due to the frequency-dependent fiber mode area, which is particularly important in quantitatively modeling the self-frequency shift experienced by the rogue solitons [20].

Rogue wave dynamics would be expected whenever SC generation is induced from an initial stage of modulation instability, and we therefore consider picosecond pulse excitation under similar conditions to previous numerical studies [15]. Specifically, we model 5 ps FWHM gaussian pulses propagating in 20 m of photonic crystal fiber with zero dispersion at 1055 nm. The dispersion coefficients at the 1060 nm pump wavelength are: $\beta_2 = -4.10 \times 10^{-1}$ ps$^2$ km$^{-1}$, $\beta_3 = 6.87 \times 10^{-2}$ ps$^3$ km$^{-1}$, $\beta_4 = -9.29 \times 10^{-3}$ ps$^4$ km$^{-1}$, $\beta_5 = 2.45 \times 10^{-7}$ ps$^5$ km$^{-1}$, $\beta_6 = -9.79 \times 10^{-10}$ ps$^6$ km$^{-1}$, $\beta_7 = 3.95 \times 10^{-12}$ ps$^7$ km$^{-1}$, $\beta_8 = -1.12 \times 10^{-14}$ ps$^8$ km$^{-1}$, $\beta_9 = 1.90 \times 10^{-17}$ ps$^9$ km$^{-1}$, $\beta_{10} = -1.51 \times 10^{-20}$ ps$^{10}$ km$^{-1}$. The input peak power $P_0 = 100$ W, $\gamma = 0.015$ W$^{-1}$ m$^{-1}$ and $\tau_{shock} = 0.66$ fs. Noise is included on the input field through a one-photon-per mode background and through a thermal spontaneous Raman scattering source term [6]. In this regard, however, we note that spontaneous Raman noise was not found to significantly influence the rogue wave statistical behavior seen in our simulations. This is in agreement with previous studies that have shown that the dominant cause of output SC instability is the nonlinear amplification of input pulse noise [7].

The general features of rogue soliton generation are shown in Fig. 1. Here, Fig. 1 (a) superposes the output spectra (gray traces) from an ensemble of 1000 simulations with different noise seeds, as well as the calculated mean (black line). The mean spectrum extends from 925–1240 nm at the −20dB level. The expanded view of the long wavelength edge in Fig. 1(b) allows us to clearly see the small number of rogue soliton events associated with a greatly increased red shift, and these particular events can be isolated using the technique developed in Ref. 15. Specifically, for each SC in the ensemble, a spectral filter selects components above a particular wavelength on the long wavelength edge, and Fourier transformation then yields a series of ultrashort pulses of varying power depending on the position of the filter relative to the SC spectral structure. The frequency distribution of the pulse peak power then readily reveals the presence of statistically-rare high peak-power rogue solitons that have been fully captured because of their extreme shifts to longer wavelengths.

Fig. 1(c) shows the histogram of the data using a filter at 1210 nm. The frequency distribution is clearly very skewed and the fraction of high power rogue solitons is extremely small. The statistically-rare nature of the rogue solitons can be seen more clearly on the log-log representation in the inset and, indeed, only 1 realization in the 1000 run ensemble has a peak power exceeding 1 kW. Although a full treatment of the statistical properties of the rogue solitons is outside the scope of this paper, our preliminary analysis suggests that the histogram is well-fitted by a Weibull distribution, a class of "extreme value" probability density function that is commonly used to analyze events associated with large deviations from the mean and median [21]. This fit is shown as the solid line in Fig. 1(c) [22].

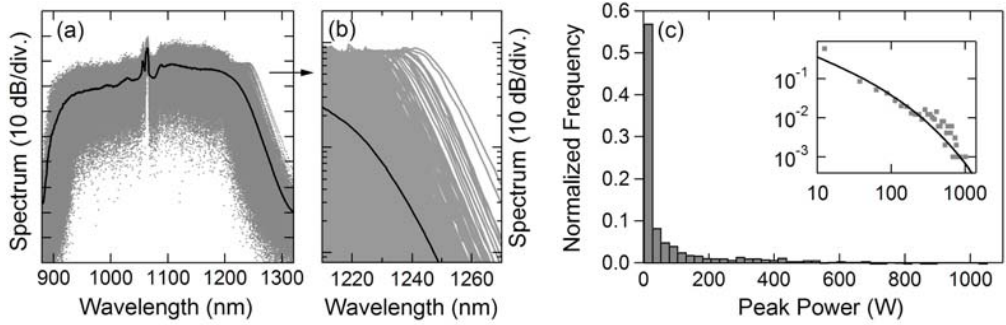

Fig. 1 (a) Results showing 1000 individual spectra (gray curves). The mean spectrum is shown as the solid black line. (b) Expanded view above 1210 nm. (c) Histogram of the peak power frequency distribution using 25 W bins. We plot normalized frequency such that bar height represents the proportion of data in each bin. The inset plots the results on a log-log scale, and also shows the fitted Weibull distribution (solid line) [22].

Additional key features of the rogue soliton dynamics are shown in Fig. 2, where we compare the spectral and temporal evolution of two selected realizations in the ensemble. In particular, Fig. 2(a) illustrates the evolution of a "rogue" event associated with the generation of a 900 W peak power soliton centered around 1240 nm, whereas Fig. 2(b) shows the evolution for a case where the output spectrum is closer to the distribution median, and there is little spectral energy above 1210 nm.

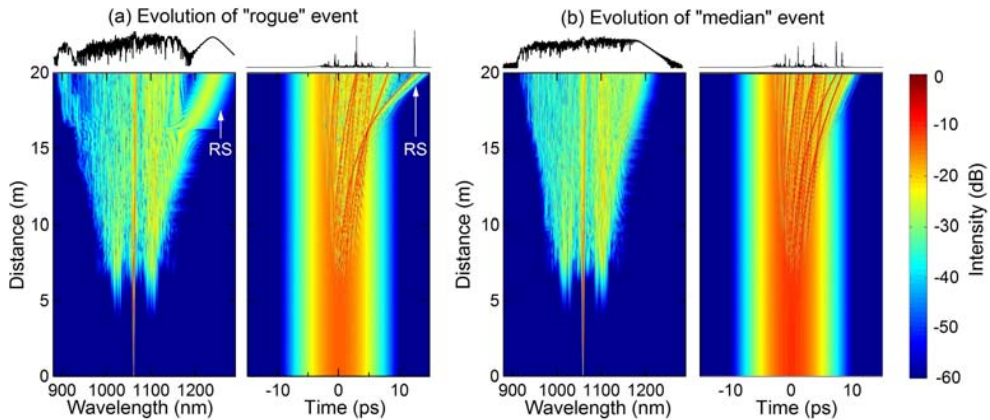

Fig. 2. Density plots of spectral and temporal evolution for: (a) a rare event leading to a rogue soliton (RS) and (b) a result close to the distribution median.

In both cases, however, the initial evolution is similar, with the growth of characteristic MI sidebands about the pump, and the development of a corresponding temporal modulation on the pulse envelope. Nonetheless, differences in the way in which the MI is seeded from the initial random noise leads to significant variation in the spectral and temporal evolution trajectories after a propagation distance of around 10 m. It is during this phase of the propagation that the modulated pulse envelope breaks up into individual soliton pulses, but we can see clearly how the rogue event is associated with the ejection of one particular soliton that propagates with significantly different group velocity, undergoes a much greater Raman frequency shift and clearly separates from other components of the evolving SC field in both the time and frequency domains. In fact, although we show only two particular results in this figure, a full analysis of the ensemble shows that the distinct spectral and temporal evolution trajectory shown in Fig. 2(a) is a characteristic feature of optical rogue soliton generation [23].

**3. Harnessing the rogue wave dynamics**

Based on this discussion of the general features of optical rogue wave generation, we now consider ways in which the underlying dynamics can be actively manipulated. Firstly, the central role played by MI in seeding the spectral broadening suggests that modifying the input pulse initial conditions will influence the rogue wave development. Indeed, a correlation between the rogue wave amplitude and a localized noise burst on the pulse leading edge has already been numerically demonstrated [15]. Here, however, we consider inducing rogue waves using more practical techniques that have been previously used in many successful experiments studying induced MI processes at THz repetition rates [24]. Specifically, we consider exciting the SC generation process using a dual frequency pumping set-up that induces a low amplitude THz modulation across the full extent of the temporal envelope.

Such an envelope modulation can be conveniently implemented experimentally by mixing a pulse with a frequency-shifted replica using wavelength conversion techniques [25].

Numerical results showing the effect of such an induced modulation on the rogue wave dynamics are shown in Fig. 3. Here, a small intensity modulation of 4% is imposed on a gaussian pulse envelope as described above, and simulations are performed for a modulation frequency varying over the range 0-20 THz spanning the MI gain bandwidth. Note that the MI gain here is calculated including the contribution of the Raman susceptibility [26]. To isolate the effect of the induced modulation, no random noise sources are included, but all other parameters are as described above. The figure shows a density plot of the output spectra obtained as a function of frequency over the MI gain profile, as well as spectral profiles for selected modulation frequencies as shown.

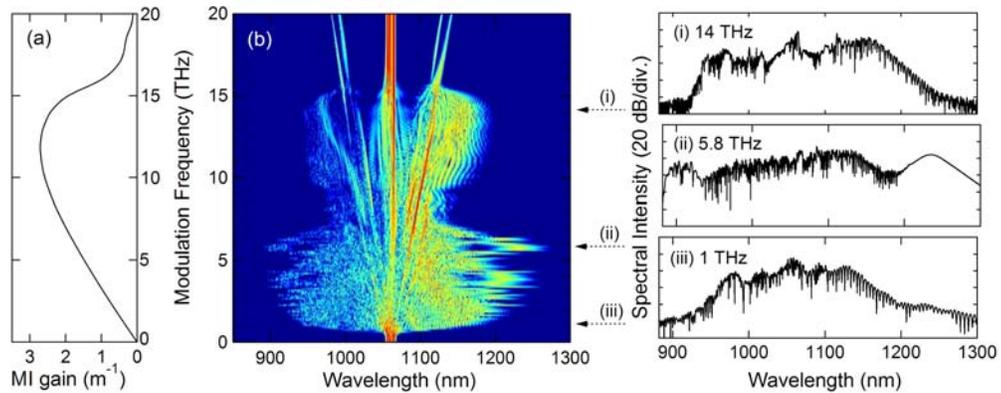

Fig. 3. (a) MI gain curve and (b) density plot of output spectra with a 4% envelope modulation at the frequency indicated. Subfigures (i)-(iii) show spectra at the frequencies indicated.

These results clearly show that the frequency of the induced modulation on the input pulse envelope plays a highly significant role in determining the output spectral structure. In fact, because the propagation length extends over a length beyond the initial phase of MI sideband growth, the final spectral structure is determined by the complex interaction between the initial MI development and the subsequent soliton dynamics. As a result, the spectral broadening at the fiber output does not correlate in a straightforward manner with the calculated MI gain curve. Although detailed studies of these propagation dynamics in the presence of induced modulation will be presented elsewhere, we can nonetheless identify a particular frequency range around 6 THz where the modulation leads to dramatically-enhanced spectral broadening, and the clear separation of an isolated Raman soliton peak. The maximum Raman soliton frequency shift is observed at a modulation frequency of 5.8 THz, and the spectrum in this case is shown explicitly in subfigure (ii) in the right panel.

Significantly, additional simulations where the modulation is imposed in the presence of input pulse noise confirm that an enhanced spectral broadening signature is still observed under realistic conditions. In particular, results from an ensemble of 1000 simulations with both a 5.8 THz modulation and broadband random noise at the quantum level show that the induced modulation acts to stimulate a dramatic increase in the number of generated rogue waves. Fig. 4 (a) shows the results obtained, where we see both an increase in the mean spectral broadening (905-1260 nm at the -20 dB level) and an increased number of rogue solitons. When compared to Fig. 1(c), the associated histogram is clearly significantly more skewed to the generation of higher-peak power events, and quantitative analysis shows that for these results, 1/100 of filtered pulses above 1210 nm have a peak power > 1 kW. This represents an order of magnitude increase compared to the case without an induced envelope modulation. Note that aside from the induced modulation, all other parameters were the same as in Fig. 1.

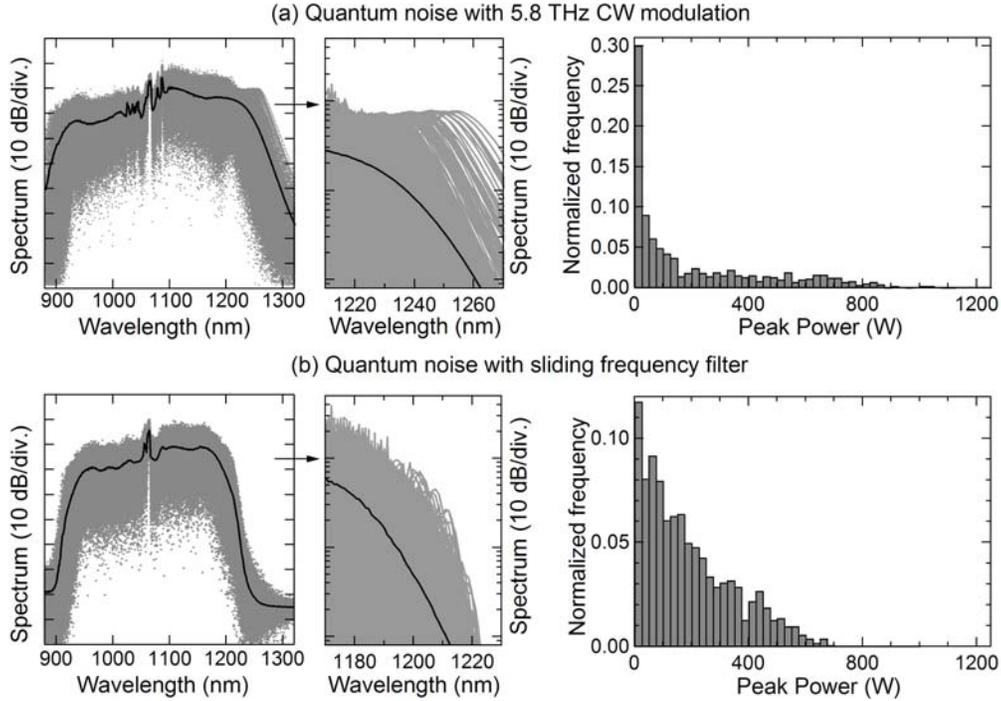

Fig. 4. Results for (a) an induced modulation at 5.8 THz and (b) a sliding frequency filter. The left panels show individual spectra (gray curves) and the corresponding mean spectrum (solid line) and the middle panels shows expanded views of the long wavelength edge. The right panels show the normalized frequency distribution of the peak power after spectral filtering.

In contrast to controlling the dynamics directly through applied modulation on the input pulse, we now consider an alternative approach that modifies the rogue wave generation through spectral filtering. In particular, the significant differences in the frequency domain trajectories of "rogue" and "median" events seen in Fig. 2 suggests a very straightforward approach to suppressing extreme rogue wave frequency shifts. Specifically, we have found that the use of a sliding frequency filter can be used to effectively attenuate these rogue soliton trajectories whilst minimizing the associated overall reduction in SC bandwidth and energy. In practice, such filtering could be implemented using long period grating technology [27, 28] and, to anticipate how this may be applied experimentally, we have carried out simulations under the same conditions as in Fig. 1 but with a series of discrete filters at 1 m intervals. To filter out rogue wave trajectories while minimizing the effect on the overall bandwidth, the filters are placed only over the propagation range 15-20 m where the differences between rogue and median trajectories are most apparent. The filters used in the simulations have 20 nm bandwidth, introduce 20 dB loss (in practice this would be reflection), and have a sliding central wavelength that varies from 1170 nm at 15 m to 1230 nm at 20 m.

Results from an ensemble of 1000 simulations in the presence of filtering are shown in Fig. 4(b), and a qualitative comparison with the results in Fig. 1 clearly demonstrates that this approach effectively reduces the generation of extreme red-shifted rogue solitons. More quantitatively, isolating potential rogue soliton pulses on the long-wavelength side (using a spectral filter at 1170 nm) yields a significantly different frequency distribution to that obtained in the absence of filtering, and no soliton pulses with a peak power > 1 kW are observed. Significantly, this is achieved whilst still maintaining reasonable spectral broadening (940-1200 nm at the -20 dB level) and whilst only decreasing the output energy of the SC by ~5%.

## 4. Discussion and outlook

The study reported here has been motivated by the recent experimental observation of optical rogue waves, statistically rare soliton pulses generated on the long wavelength edge of a broadband SC spectrum. Our numerical simulations have provided insight into the distinctive features of these rogue soliton events, and we have considered two specific ways by which their dynamics and statistical properties can be harnessed and controlled using available experimental techniques.

Perhaps the simplest technique is the use of a sliding frequency filter to attenuate extreme red-shifting rogue wave trajectories in the spectral domain. Simulations have shown that discrete filtering can efficiently remove the rogue soliton contribution to the output spectrum with only minor reduction in SC bandwidth and energy. An area where such filtering may prove especially important is for SC generation in highly nonlinear glass fibers [29], where suppressing the highest intensity soliton pulses could allow SC generation at higher average powers without deleterious photodarkening.

Modulation of the input pulse profile has also been studied as a means of modifying the propagation dynamics in a more direct way. An applied modulation at 5.8 THz has been shown to lead to an order of magnitude increase in the generation rate of extreme-red shifted solitons and such modulation could be applied in practice by mixing a pump with a frequency-shifted replica using well-established wavelength conversion techniques. In this regard, although we have studied the effect of modulation at only one particular frequency with the goal of enhancing rogue wave generation, we anticipate that varying the modulation parameters over a wider range may well allow the properties of SC spectra to be tailored in a more general way.